\documentclass[aps,pra,twocolumn,showpacs,groupedaddress]{revtex4-1}
\usepackage{pifont}
\usepackage{amsmath}
\usepackage{graphicx}
\usepackage{amssymb}
\usepackage{amsfonts}
\usepackage{subfigure}
\usepackage{booktabs}
\usepackage{setspace}
\usepackage{threeparttable}
\usepackage{enumerate}
\makeatletter

\newcommand{\Rmnum}[1]{\expandafter\@slowromancap\romannumeral #1@}
\makeatother

\begin{document}

\title{Collective modes of spin-orbit coupled Fermi gases in the repulsive regime}
\author{Shang-Shun Zhang$^1$}
\author{Xiao-Lu Yu$^{1}$}
\author{Jinwu Ye$^{2,3}$}
\author{Wu-Ming Liu$^1$}
\affiliation{ $^{1}$Beijing National Laboratory for Condensed Matter Physics, Institute of
Physics, Chinese Academy of Sciences, Beijing 100190, China \\
$^{2}$ Department of Physics and Astronomy, Mississippi State University, MS 39762, USA \\
$^{3}$ Key Laboratory of Terahertz Optoelectronics, Ministry of Education, Department of Physics, Capital Normal University, Beijing 100048, China}
\date{\today}

\begin{abstract}
We investigate the collective modes in the spin-orbit coupled Fermi gas with repulsive $s$-wave interaction. The interplay between spin-orbit coupling and atom-atom interactions plays the crucial role in the collective behaviors of Fermi gas. In contrast with ordinary Fermi liquid, spin-orbit coupled Fermi gas has strongly correlated spin and density excitations. Within the scheme of random phase approximation, we classify collective modes based on the symmetry group and determine their properties via the poles of corresponding correlation functions. Besides, the particle-hole continuum is obtained, where the imaginary part of these correlation functions become non-vanishing. We also propose an experimental protocol for detecting these collective modes and discuss corresponding experimental signatures in the ultracold Fermi gases experiment.
\end{abstract}

\maketitle

\section{Introduction}
A great deal of attentions have been focused on the spin-orbit coupling (SOC) because of its fundamental interests in condensed matter systems \cite{Kato1,Wunderlich,Bernevig,Konig,Kane,Bernevig2} and important applications in spintronic device \cite{Koralek}. In recent years, a wide range of atomic physics and quantum optics technology provides unprecedented manipulation of a variety of intriguing quantum phenomena, therefore it seems to provide an ideal platform to study the effects of SOC in ultracold atomic systems. The experimental studies on this topic have made great breakthrough. Based on the Berry phase effect \cite{Berry1984} and its non-Abelian generalization \cite{Wilczek1984}, Spielman's group in NIST has successfully generated a synthetic external Abelian or non-Abelian gauge potential coupled to neutral atoms \cite{NIST1,NIST2,NIST3,NIST4}. Recently, the SOC Fermi gas has been first engineered in weakly repulsive $^{40}$K \cite{JZhang} or $^{6}$Li \cite{Zwierlein} atomic gases. Realization of SOC in quantum gases will open a whole new avenue in cold atom physics.

Motivated by these recent experimental progresses of ultracold Fermi gases, we consider the two dimensional (2D) Fermi gas with SOC in the repulsive regime. The repulsive atom-atom interaction can be engineered in the upper branch of energy spectrum, where there are uncondensed Fermi gas in the absence of molecule formation \cite{TLHo}. The repulsive Fermi gas is stable when it is far away from the resonant regime, which has been successfully reached in experiment \cite{GBJo}. In our previous work, we have studied low energy single particle excitations and calculated the Fermi liquid parameters such as the quasi-particle lifetime, renormalization factor and the effective mass in the repulsive regime \cite{XLYu}.

In this paper, we investigate the collective modes of two dimensional Fermi gas with SOC in the repulsive regime. The research of the low energy collective modes in degenerated quantum gases yields a wealth of insights into the properties of ultracold atomic systems. A lot of previous studies have been devoted to the collective behaviors in various symmetry broken phases of ultracold atomic systems, which include the degenerate gas at the Bose-Einstein condensation (BEC) to Bardeen-Cooper-Schrieffer (BCS) crossover \cite{HHu,Wright,Bartenstein,Altmeyer}, Fermi gas in the unitarity limit \cite{Riedl,Kinast1,Kinast2,Stringari}, imbalanced Fermi gas \cite{Nascimbene}, and BEC in the presence of SOC \cite{JYZhang,HZhai}. By contrast, we focus on the normal state regime of the SOC Fermi gas in this work. The interplay between SOC and $s$-wave atom-atom interactions plays a crucial role in our investigation. Compared with ordinary Fermi gas, SOC Fermi gas has strongly correlated spin and density excitations. Within the scheme of random phase approximation (RPA), we classify all the collective modes and determine their properties via the poles of corresponding RPA correlation functions.

In contrast with previous works in solid state systems \cite{Tanmoy,HaoChen,Ryan2,Gammon,Badalyan,XLQi,book:Gabriele,Agarwal,Robert}, the consideration of $s$-wave interaction in ultracold atomic systems instead of the Coulomb interaction leads to qualitatively different collective behaviors. The reasons for this are twofold: First, the force range of the $s$-wave interaction is short and the interaction vertex is independent to the momentum transfer. Second, the $s$-wave interaction is spin-dependent in ultracold Fermi gas, which can be decomposed into the density and spin channels \cite{book:Mahan}. The collective modes of this system are grouped into two categories: (i) one branch of gapless mode, namely zero sound, which is an oscillation of density coupled with the transverse spin oscillation; (ii) three branches of gapped modes. We expect our microscopic calculations of the collective modes would have immediate applications to the SOC Fermi gas in the upper branch of the energy spectrum.

The paper is organized as follows. The model building of SOC Fermi gas with repulsive $s$-wave interaction is described in Sec. II, where all the microscopic parameters and the helicity eigenstates are explained. Furthermore, the collective modes are classified based on the symmetric property. In Sec. III, we develop a general approach to calculate a series of RPA correlations functions. In Sec. IV, we investigate the solutions of matrix forms of RPA equation both analytically and numerically. Finally, we propose an experiment protocol to detect collective modes in Fermi gas with SOC on the upper branch of energy spectrum and estimate the corresponding experimental signatures in Sec. V.

\section{The SOC Fermi gas with repulsive $S$-wave scattering}
\begin{figure}[t]
\includegraphics[width=3in]{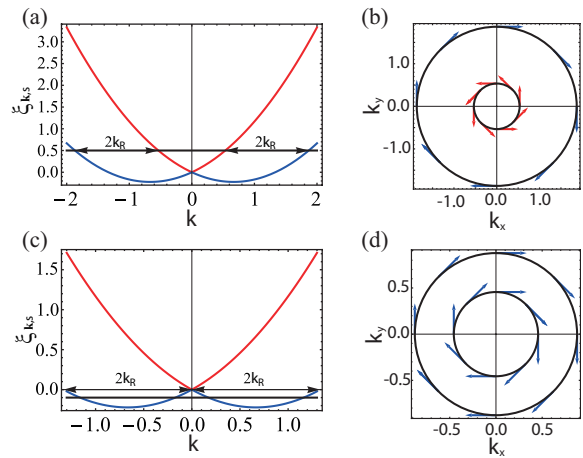}
\caption{(Color online) (a) and (c) plot the energy spectrum in presence of spin-orbit coupling with different fillings. The thick black horizonal line denotes the level of chemical potential. (b) and (d) show the Fermi surfaces and the associated spin textures corresponding to (a) and (c) respectively.}
\label{spectrum}
\end{figure}

We consider the collective modes of a 2D spin-1/2 ultracold repulsive Fermi gas with Rashba SOC, described by the model Hamiltonian,
\begin{eqnarray}  \label{eq:H}
\mathcal{H}=\! \! \sum\limits_{\mathbf{k},\alpha,\beta}\!\! c^{\dag}_{%
\mathbf{k},\alpha}\! h_{\alpha \beta} c_{\mathbf{k},\beta} \! +\!2g\!\!
\sum\limits_{\mathbf{k},\mathbf{p},\mathbf{q}}c_{\mathbf{k}+\mathbf{q}%
,\uparrow}^{\dag} c_{\mathbf{p}-\mathbf{q},\downarrow }^{\dag} c_{\mathbf{p}%
,\downarrow } c_{\mathbf{k},\uparrow } ,
\end{eqnarray}
where $\alpha$ and $\beta$ are the spin indexes. The first term is the non-interacting part, and $h$ is the single particle Hamiltonian with SOC \cite{Juzeliunas}
\begin{equation}  \label{eq:H0}
h=\frac{\mathbf{k}^{2}}{2m}+\lambda( \hat{\mathbf{z}}\times \mathbf{
k}) \cdot \mbox{\boldmath$\sigma$} -\mu,
\end{equation}
where $\mu=k_{F}^{2}/2m$ is the chemical potential and $k_F$ is the Fermi momentum in the absence of SOC, $\lambda $ represents the strength of SOC. Hereafter, $\hbar$ is taken as $1$. The eigenstates of single particle Hamiltonian $h$ can be obtained as follows
\begin{equation}  \label{eigenstates}
|\mathbf{k},\pm 1 \rangle =\frac{1}{\sqrt{2}}\left(
\begin{array}{c}
1 \\
\pm ie^{i\phi (\mathbf{k}) }%
\end{array}
\right),
\end{equation}
where $\phi(\mathbf{k})$ is the azimuthal angle of momentum $\mathbf{k}$ and the helicity $\pm 1$ represents that the in-plane spin polarization is right-handed or left-handed with respect to the momentum. The dispersion relations for two helicity branches are $\xi _{ \mathbf{k},\pm}=(\mathbf{k}^{2}\pm 2k_{R}|\mathbf{k}|-k_{F}^{2})/2m$, where $k_{R}=m\lambda$ corresponds to the recoil momentum in experiments \cite{JZhang,Zwierlein}. The Fermi momentum and the recoil momentum provide two scales for the 2D Fermi gas with SOC, and the dimensionless ratio $\gamma=k_{R}/k_{F}$ denotes the significance of SOC. The Fermi surfaces are given by $\xi _{ \mathbf{k}s}=0$, which give rise to two circles in the momentum space with two Fermi momenta as $k_{s}=\kappa k_F -sk_{R}$, where $\kappa =\sqrt{1+\gamma ^{2}}$. We plot the energy spectrum and the Fermi surfaces with the associated spin textures in Fig. \ref{spectrum} for two different chemical potentials respectively. We note that the outer and inner Fermi surfaces shown in Fig. \ref{spectrum} (b) have opposite helicities, therefore one has Berry phase $\pi$, another $-\pi$. While the outer and inner Fermi surfaces shown in Fig. \ref{spectrum} (d) have the same helicity and Berry phase $\pi$. In the present work, we only consider the case shown in Fig. \ref{spectrum} (a), where the two Fermi surfaces correspond to different helicity branches.

The second term in Eq. (\ref{eq:H}) represents the $s$-wave interaction in ultracold atomic gases. The low energy interaction among ultracold atoms is universally determined by the scattering length $a_s$ \cite{Mott,Messiah,Bratten}. For quasi-2D system, which can be realized through a strong confinement in the $\mathbf{\hat{z}}$ direction perpendicular to the 2D plane, the effective $s$-wave scattering strength is determined by $2g=4\pi N a_s/3 \sqrt{2\pi} m \zeta_z$, where $N$ is the total atom number, $m$ is the mass for atom, and $\zeta_z=\sqrt{1/m\omega_z}$ is the confinement length of the atomic gases along $\mathbf{\hat{z}}$ direction with $\omega_z$ as the trap frequency of the confinement potential. The significance of $s$-wave interaction in ultracold atoms could be characterized by the dimensionless ratio of the average interaction energy and kinetic energy $\varepsilon_{int}/\varepsilon_{kin}$, i.e., $mg=2\pi N a_s/3 \sqrt{2\pi}\zeta_z$. With the technique of Feshbach resonance, the strength of the repulsive interaction can be tuned within a wide range \cite{Kohler,Chin,Inouye,Courteille,Roberts,Vuletic}.

The dynamical response of SOC Fermi gas can be described via the density and spin susceptibility
\begin{eqnarray}
\chi ^{\mu \nu }\left( \mathbf{q},i\omega _{m}\right) &=&\sum\limits_{\mathbf{k,p}}\sum\limits_{\alpha \beta \gamma \delta }\int_{0}^{\beta }d\tau
e^{i\omega _{m}\tau }\langle \mathcal{T}_{\tau }c_{\mathbf{k}\alpha }^{\dagger}\sigma
_{\alpha \beta }^{\mu }c_{\mathbf{k+q}\beta }(\tau )  \notag \\
&& \times c_{\mathbf{p+q}\gamma }^{\dagger}\sigma _{\gamma \delta }^{\nu }c_{\mathbf{p}
\delta }\left( 0\right) \rangle,
\end{eqnarray}
where $\sigma ^{\mu }=(\sigma^0,\mbox{\boldmath$\sigma$})$, $\sigma^0$ is the $2\times 2$ unit matrix and $\mbox{\boldmath$\sigma$}$ are the Pauli matrixes. We could study the collective modes through the poles of the dynamical response function to external density and spin perturbations. The SOC is isotropic in the 2D $x$-$y$ plane. The density and spin susceptibility $\chi^{\mu\nu}$ is invariant under the simultaneous rotation of the momentum $\mathbf{q}$ and the spin $\mathbf{s}=\frac{1}{2}\mbox{\boldmath$\sigma$}$ around the $\mathbf{\hat{z}}$ axis, namely the rotation group: $SO(2)_{\mathcal{D}}$, where the subscript $\mathcal{D}$ represents the combined operation for the momentum and spin space.

Furthermore, considering the Rashba-type SOC and the rotation invariance of Fermi surface around the $\mathbf{\hat{z}}$ axis, the density excitations are coupled with the transverse components of the spin excitations $s^{T}\equiv (\mathbf{\hat{z}\times {\hat{q})\cdot \mathbf{s}}}$ intrinsically, while both of them are decoupled from the longitudinal component $s^{L}\equiv \mathbf{\hat{q}}\cdot \mathbf{s}$ and perpendicular component $s^{Z}\equiv \mathbf{ \hat{z}}\cdot \mathbf{s}$ \cite{Bernevig3}. Therefore, it is convenient to study the collective modes under the helical representation, which is defined as $\{n,s^{T},s^{L},s^{Z}\}$ with $n$ as the density component. Under the helical representation, the $4\times 4$ susceptibility matrix decomposes into two $2\times 2$ matrices. One is in the subspace of $n$-$s^T$, and the other is in the subspace of $s^L$-$s^Z$. Based on this decomposition of the susceptibility, the collective modes are grouped into two categories: one with the density and transverse spin excitations $s^{T}$, and the others with the perpendicular and longitudinal spin excitations $s^{Z},s^{L}$. Without loss of generality, we can assume $\mathbf{q}=q\mathbf{{e_{x}}}$ in the following. Thus we have
\begin{equation}
s^{T}=s^{y},s^{L}=s^{x},s^{Z}=s^{z}.
\end{equation}
In the following sections, we will work in this representation to calculate the density and spin susceptibility and investigate the collective modes of SOC Fermi gas.

\section{density and spin correlation functions}

\subsection{Feynman Rules and Susceptibilities}

\begin{figure}[t]
\begin{center}
\includegraphics[width=3in]{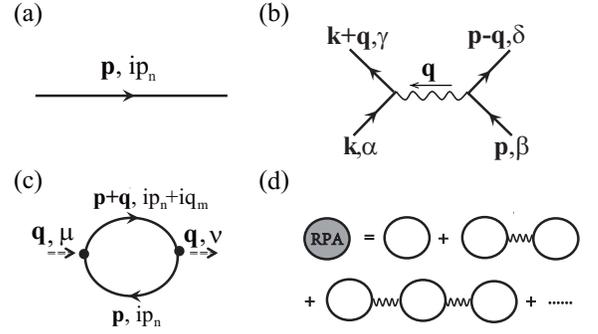}\hspace{0.5cm}
\end{center}
\caption{(Color online) (a) The free Green's function with SOC: $-G^0(\mathbf{p} ,ip_n)$. (b) The interaction vertex $-g\left( \protect\sigma ^{\protect\mu }\right)^{\protect\beta \protect\gamma }\left( \protect\sigma _{\protect\mu }\right) ^{\protect\alpha \protect\delta } $. (c) One bubble
diagram: the bare susceptibility $\protect\chi^{\protect\mu \protect\nu}(\mathbf{q},iq_m)$. (d) The sum of the bubble diagrams gives the RPA susceptibility $\protect\chi_{RPA}^{\protect\mu \protect\nu }(\mathbf{q},iq_m)$. }
\label{Feynman}
\end{figure}

The particle line shown in Fig. \ref{Feynman} (a) represents the Green's function with SOC in the Matsubara formalism
\begin{equation}
G_{\alpha \beta }^{0}\left( \mathbf{k},ik_{n}\right) =\sum\limits_{s=\pm 1}\frac{
(P_{s})_{\alpha \beta }}{ik_{n}-\xi _{\mathbf{k},s}},
\end{equation}
where $i k_n=(2 n+1) \pi k_B T$ is the fermionic Matsubara frequency and $P_{s}$ is the projection operator to the helicity eigenstates: $P_{s}=[1+s(\mathbf{\hat{z}\times \hat{k}})\cdot \mbox{\boldmath$\sigma$}]/2$. It is helpful to apply the Fierz identity $2\epsilon ^{\delta \gamma }\epsilon^{\alpha \beta }=\left( \sigma ^{\mu }\right) ^{\beta \gamma }\left( \sigma_{\mu }\right) ^{\alpha \delta }$ to the $s$-wave interaction vertex, where $\epsilon^{\alpha \beta}$ is the $2\times 2$ antisymmetric matrix, $\sigma ^{\mu }=(\sigma ^{0},\mbox{\boldmath$\sigma$})$ and $\sigma _{\mu}=(\sigma ^{0},-\mbox{\boldmath$\sigma$})$. The interacting part can be rewritten as
\begin{eqnarray}
V&=\frac{1}{2!2!}\sum\limits_{\alpha \beta \gamma \delta
}\sum\limits_{\mathbf{k,p,q}}V_{\alpha \delta ,\beta \gamma }
c_{\delta }^{\dagger}\left( \mathbf{k+q}\right) c_{\gamma }^{\dagger}\left( \mathbf{p-q}\right)  \notag
\\
& c_{\beta }\left( \mathbf{p}\right) c_{\alpha }\left( \mathbf{k}\right) ,
\end{eqnarray}
where $V_{\alpha \delta ,\beta \gamma }=g\left( \sigma ^{\mu}\right)_{\beta \gamma }\left( \sigma _{\mu }\right)_{\alpha \delta }$. The Feynman rule for the interaction vertex is shown in Fig. \ref{Feynman} (b). In fact, the interaction vertex considered here includes two cases: $(\mathbf{k} \uparrow, \mathbf{p} \downarrow) \rightarrow (\mathbf{k^{\prime}}\uparrow, \mathbf{p^{\prime}}\downarrow)$ and $(\mathbf{k}\uparrow, \mathbf{p} \downarrow)\rightarrow (\mathbf{k^{\prime}}\downarrow, \mathbf{p^{\prime}} \uparrow)$, which corresponds to direct and exchange interaction respectively \cite{Fetter,Wen}.

The collective modes and particle-hole pairs are two fundamental types of excitations
of the SOC Fermi gas. The dispersions of collective modes are determined by the poles of the density and spin
susceptibility. With these Feynman rules defined above, the RPA susceptibility can be evaluated as (see Fig. \ref{Feynman} (d))
\begin{equation}  \label{eq:chiRPA}
\chi _{RPA}^{\mu \nu }\left( \mathbf{q},iq_{m}\right) =\chi^{\mu \rho }\left( \mathbf{q},iq_{m}\right)(\frac{1}{1+g\eta \chi\left( \mathbf{q},iq_{m}\right)})^{\rho \nu },
\end{equation}
where
\begin{equation}
\eta =\text{diag}\{1,-1,-1,-1\},
\end{equation}
and $\chi$ is the bare susceptibility for noninteracting Hamiltonian.

\begin{figure}[!t]
\begin{center}
\includegraphics[width=2.52in]{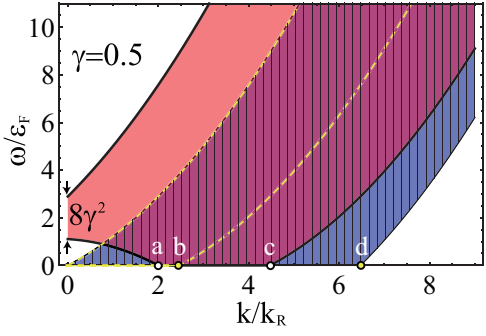}\hspace{0.5cm}
\end{center}
\caption{(Color online) Particle-hole continuum of SOC Fermi gas for $\gamma=0.5$. The red region surrounded by the thick black lines represents the interband particle-hole continuum. The region surrounded by the dashed yellow lines represents continuum of intraband particle-hole excitations with helicity $+1$, the blue region filled with vertical lines, with helicity $-1$. The points $a$, $b$, $c$ and $d$ correspond to the momenta $2k_R$, $2k_{+1}$, $2 \kappa k_F$ and $2 k_{-1}$ respectively, where the static susceptibility function exhibits singular behaviors \cite{GHChen}.
}
\label{figure:phc}
\end{figure}

In Matsubara formalism, the bare susceptibility $\chi^{\mu \nu }(%
\mathbf{q},iq_{m})$ is given by the one loop diagram as shown in Fig. %
\ref{Feynman} (c)%
\begin{eqnarray}  \label{chi0}
\chi^{\mu \nu }\left( \mathbf{q},iq_{m}\right) &=-k_{B}T\sum\limits_{%
\mathbf{k},ik_{n}} \text{Tr} [ G^{0}\left( \mathbf{k+q}/2,ik_{n}\right) \sigma
^{\mu }  \notag \\
& \times G^{0}\left( \mathbf{k-q}/2,ik_{n}-iq_{m}\right) \sigma^{\nu }],
\end{eqnarray}%
where $iq_{m}=2m\pi k_{B}T$ and $ik_{n}=(2n+1)\pi k_{B}T$ are the bosonic
and fermionic Matsubara frequencies, respectively. The Green's function $G^{0}\left( \mathbf{k},ik_{n}\right) $ includes a momentum-dependent projection operator $P_s$, and the trace in Eq. (\ref{chi0}) gives rise to an overlap factor as
\begin{equation}  \label{eq:factor}
F_{sr}^{\mu \nu }=\text{Tr}\left[ P_{s}\left( \mathbf{k+q}/2\right) \sigma^{\mu
}P_{r}\left( \mathbf{k-q}/2\right) \sigma^{\nu }\right],
\end{equation}%
where $\mu ,\nu = 0,1,2,3$ represent the density and spin components
in $\mathbf{\hat{x}},\mathbf{\hat{y}},\mathbf{\hat{z}}$ directions, and $s,r$ are helicity indexes. After summing over the fermionic Matsubara frequency $ik_{n}$, and performing the
analytical continuation $iq_{m}\rightarrow \omega +i0^{+}$, we have
\begin{equation}  \label{susceptibility}
\chi^{\mu \nu }\left( \mathbf{q},\omega \right) =-\sum\limits_{\mathbf{%
k,}s,r}F_{sr}^{\mu \nu }\frac{f\left( \xi _{\mathbf{k-q/}2,s}\right)
-f\left( \xi _{\mathbf{k+q/}2,r}\right) }{\xi _{\mathbf{k-q/}2,s}-\xi _{%
\mathbf{k}+\mathbf{q}/2,r}+\omega +i0^{+}}.
\end{equation}
At zero temperature, the Fermi occupation function is $f\left( \xi \right)
=\Theta (-\xi )$. The numerator in the Eq. (\ref{susceptibility}) is non-zero only if
\begin{equation}  \label{condition1}
\xi _{\mathbf{k-q/}2,s}>0, \ \ \xi _{\mathbf{k+q/}2,r}<0,
\end{equation}
or
\begin{equation}  \label{condition2}
\xi _{\mathbf{k-q/}2,s}<0, \ \ \xi _{\mathbf{k+q/}2,r}>0,
\end{equation}
which represent the conditions for particle-hole excitations. The $\xi _{\mathbf{k-q/}2,s}-\xi _{\mathbf{k}+\mathbf{q}/2,r}$ in the denominator is the corresponding particle-hole exciton energy, which includes the contribution from intraband and interband and forms a continuum as shown in Fig. \ref{figure:phc}. Accordingly, the contributions to the susceptibility come from two aspects: the intraband $r=s $ and the interband $r=-s$. When the frequency and momentum of the susceptibility $\omega, \mathbf{q}$ fall in the particle-hole continuum (see Fig. \ref{figure:phc}), the integration in Eq. (\ref{susceptibility}) will develop a non-zero imaginary part, which corresponds to the damping of the collective excitations. The details about this integration will be further discussed in the following sections and Appendix A. The static density-density susceptibility $\chi^{00}(\mathbf{q})$ exhibits singular behaviors at $|\mathbf{q}|=2k_R$ and $|\mathbf{q}|=2\kappa k_F$. In addition, there are weak anomalies at $|\mathbf{q}|=2k_{+1}$ and $|\mathbf{q}|=2k_{-1}$ caused by the intraband virtual transitions \cite{GHChen}. These specific momenta are shown in Fig. \ref{figure:phc}, which appear as edges of the intra- and inter-band particle-hole excitation continuum.

We want to make some general observations of the overlap factor $F_{sr}^{\mu,\nu} $ and bare susceptibility $\chi$ as following: (i) $F_{sr}^{\dagger}\left( \mathbf{q}\right) =F_{sr}\left( \mathbf{q}\right) $. (ii) $F_{sr}^{\mu \nu }\left( \mathbf{q}\right) =F_{rs}^{\nu \mu }\left( -\mathbf{q }\right)$. (iii) $\chi\left( \mathbf{q},\omega \right) ^{\dagger}=\chi\left( - \mathbf{q},-\omega \right) $. In the long wave length limit $\left\vert \mathbf{q}\right\vert \rightarrow 0 $, the overlap factor $F^{\mu \nu }$ can be expanded relative to $\mathbf{q}$ up to $\mathcal{O}(\mathbf{q}^{2})$ approximately.  The intraband contribution is
\begin{equation}
F_{s,s}^{\mu \nu }=\left(
\begin{array}{cccc}
1 & -s \cos \theta & s \sin{\theta} & i\frac{|\mathbf{q}|\sin{\theta}}{2|\mathbf{k}|} \\
-s \cos{\theta } & \cos^{2}{\theta } & -\frac{\sin{2\theta}}{2} &  -i\frac{s|\mathbf{q}|\sin{2\theta }}{4|\mathbf{k}|}\\
s \sin{\theta} & -\frac{\sin{2\theta}}{2} & \sin^{2}{\theta } & i\frac{s|\mathbf{q}| \sin^{2}{\theta }}{2|\mathbf{k}|} \\
-i\frac{|\mathbf{q}|\sin{\theta }}{2|\mathbf{k}|} & i\frac{s|\mathbf{q}|\sin{2\theta}}{4|\mathbf{k}|} & -i\frac{s|\mathbf{q}|\sin^{2}{\theta }}{2|\mathbf{k}|} & 0%
\end{array}%
\right) ,  \label{Fss}
\end{equation}
where $\theta $ is the azimuthal angle of $\mathbf{\hat{k}=(}\cos \theta,\sin \theta \mathbf{)}$. The interband contribution is
\begin{equation}
F_{s,-s}^{\mu \nu }=\left(
\begin{array}{cccc}
0 & \frac{s|\mathbf{q}|\sin^{2}{\theta }}{2|\mathbf{k}|} & \frac{s|\mathbf{q}|\sin{2\theta}}{4|\mathbf{k}|} & -i\frac{|\mathbf{q}| \sin{\theta}}{2 |\mathbf{k}|} \\
\frac{s|\mathbf{q}|\sin^{2}\theta}{2|\mathbf{k}|} & \sin^{2}{\theta } & \frac{\sin{2\theta}}{2} & -is \sin {\theta}\\
\frac{s|\mathbf{q}|\sin{2\theta }}{4|\mathbf{k}|} & \frac{\sin{2\theta}}{2} & \cos^{2}{\theta } & -is \cos{\theta } \\
i\frac{|\mathbf{q}| \sin{\theta}}{2 |\mathbf{k}|} & i s \sin{\theta} & is \cos{\theta } & 1
\end{array}
\right).
\end{equation}
In the following calculations, we will find that the susceptibility decomposes into two $2\times 2$ matrices, which coincides with the argument in Sec. II. B based on the symmetry property of this  system. Besides, the momentum $\mathbf{q}$ dependent terms in the overlap factor have negligible contributions to the low-$q$ (long wave length limit) properties of the collective modes, such as the sound velocity of the gapless mode and the energy gaps of the gapped modes.

\subsection{Intraband contributions ($r=s$)}

In the long wave length and low frequency limit ($q\ll k_{F}$, $\omega \ll \mu $), the expansions of the particle-hole exciton energy and occupation functions in Eq. (\ref{susceptibility}) to the leading order of $|\mathbf{q}|$ are
\begin{equation}
\xi _{\mathbf{k-\frac{q}{2}},s}-\xi _{\mathbf{k}+\mathbf{\frac{q}{2}},s}\simeq-%
\frac{k+sk_{R}}{m}\mathbf{\hat{k}}\cdot \mathbf{q},
\end{equation}%
and
\begin{equation}
f( \xi _{\mathbf{k-\frac{q}{2}},s}) -f(\xi _{\mathbf{k+\frac{q}{2}%
},s})\simeq-\delta \left( \xi _{\mathbf{k},s}\right) (\xi _{\mathbf{k}-\mathbf{%
\frac{q}{2}},s}-\xi _{\mathbf{k+\frac{q}{2}},s}).
\end{equation}
%Define the dimensionless ratio: $\nu =\omega /\epsilon_{F}$, where $\epsilon _{F}=k_{F}^{2}/2m$.
Both have corrections of $\mathcal{O}(q^{2}/k_{F}^{2})$. The intraband susceptibility reads as
\begin{equation}
\chi^{\mu \nu }_{\text{intra}}\left( \mathbf{q},\omega\right) =-\frac{m}{4\pi
^{2}\kappa }\sum\limits_{s}\frac{k_{s}}{k_F}\int d\theta F_{ss}^{\mu \nu }\frac{\cos
\theta }{y^{+}-\cos \theta },
\end{equation}
where $y=m \omega/\kappa k_F |\mathbf{q}|$ and $\theta $ is the angle between $\mathbf{k}$ and $\mathbf{q}$. We notice that the intraband susceptibility is a function of the dimensionless value $y$ in the long wave length and low frequency limit. Making use of the overlap factor given in Eq. (\ref{Fss}), the elements of $\chi\left( \mathbf{q},\omega \right) $ can be expressed in terms of the following integrals
\begin{equation}
\chi^{\text{intra}}\left( \mathbf{q},\omega \right) =-\frac{m}{\pi }\left(
\begin{array}{cccc}
I_{1} & \frac{\gamma }{\kappa }I_{2} & 0 & 0 \\
\frac{\gamma }{\kappa }I_{2} & I_{3} & 0 & 0 \\
0 & 0 & I_{5} & 0 \\
0 & 0 & 0 & 0%
\end{array}%
\right) ,  \label{intraband}
\end{equation}%
where the definitions and results of these integrations $I_{1},I_{2},I_{3}$, and $I_{5}$ are given in Appendix A.

\subsection{Interband contributions ($r=-s$)}

Similar to the intraband contribution, the energy of the interband particle-hole excitation is expanded relative to $\mathbf{q}$ to the leading order
\begin{equation}\label{eq:interband}
\xi _{\mathbf{k-\frac{q}{2}},s}-\xi _{\mathbf{k}+\mathbf{\frac{q}{2}},-s} \simeq
\frac{k}{m}\left( 2s k_R -\hat{\mathbf{k}} \cdot \mathbf{q} \right),
\end{equation}
with corrections of $\mathcal{O}(q^{2}/k_{F}^{2})$. We notice that the energy of interband particle-hole excitation obtains an additional term $\pm 2 k k_R/m$ in the presence of SOC. This term dominates the energy of particle-hole excitation at the region $q \ll k_R$, which is dramatically important for giving rise to the low-$q$ properties of the collective excitations, including the sound speed of the gapless modes and the energy gaps of the gapped modes. The difference of occupation is expanded as
\begin{eqnarray} \label{deltF}
&&f(\xi _{\mathbf{k-\frac{q}{2}},s})-f(\xi _{\mathbf{k+\frac{q}{2}}%
,-s}) \simeq - s\Theta \left( k_R -\left\vert k-\kappa k_F \right\vert \right)   \notag
\label{occupation} \\
&&+[\delta \left( k-k_{s}\right)+\delta \left(
k-k_{-s}\right)] \frac{\hat{\mathbf{k}} \cdot \mathbf{q}}{2}.
\end{eqnarray}
Since the low-$q$ behaviors of collective modes provide qualitative features of the dispersion relations, we will discuss the low-$q$ properties of $\chi^{\text{inter}}$ in the following and neglect the second term of the occupation difference in Eq. (\ref{deltF}), which is negligible for properties of the collective modes in the low-$q$ regime where $|\mathbf{q}| \ll k_R$.

By contrast with the intraband susceptibility, the interband one $\chi^{\text{inter}}$ has a well defined value in the long wave length and low frequency limit, which reads as
\begin{equation} \label{chi_inter_static}
\chi^{\text{inter}} = \frac{m}{\pi}\left(
\begin{array}{cccc}
0 & 0 & 0 & 0 \\
0 & 1/2 & 0 &0\\
0 & 0 & 1/2 &0 \\
0 & 0 & 0 & 1%
\end{array}
\right),  \mathbf{q}, \omega \rightarrow 0,
\end{equation}
with corrections of $\mathcal{O}(q,\omega)$, and it has no dependence on the way that $\mathbf{q}, \omega$ tends to zero. Combined with the intraband contribution in Eq. (\ref{intraband}), we find the total susceptibility in the static and uniform limit with $\omega/q \rightarrow 0$ reads as $\chi^{\mu\nu}=\frac{m}{\pi}\delta_{\mu\nu}$, namely the total density of states above the Dirac point of the spectrum. Applying the Stoner's criterion for itinerant ferromagnetism in the framework of RPA, i.e., $\text{det}[\chi_{\text{RPA}}^{-1}]<0$, we find a critical value of the interaction strength $g_c=\frac{\pi}{m}$, above which the SOC Fermi gas experiences a Stoner instability of ferromagnetism \cite{Stoner,Duine,LeBlanc}. Experimentally, the itinerant ferromagnetism in a Fermi gas of ultracold atoms has been observed \cite{GBJo}. In the normal state regime, we have $g<g_c$, and all the evaluations in our work are focused on this regime.

Next, we consider the properties of $\chi^{\text{inter}}$ in high-frequency regime. From Eq. (\ref{intraband}), we know that the intraband contribution in this regime vanishes. Therefore only the interband susceptibility contributes, which reads as
\begin{equation}
\chi^{\text{inter}}=\frac{m}{\pi }\sum_{s}\left(
\begin{array}{cccc}
0 & 0 & 0 & 0 \\
0 & F_s/2 & 0 &0\\
0 & 0 & F_s/2 &0 \\
0 & 0 & 0 & F_s%
\end{array}
\right) +\mathcal{O}(\frac{\mathbf{q}}{k_R}),  \label{interband12}
\end{equation}%
where $F_s=(2 \gamma-z_s \text{ln}\frac{z_s+\kappa+\gamma}{z_s+\kappa-\gamma})/4\gamma$ with $z_s=\omega/4s\gamma \epsilon_F$.

The summation of the intraband and interband contribution gives the total susceptibility
\begin{eqnarray}
\chi =\chi^{\text{intra}} +\chi^{\text{inter}}.
\end{eqnarray}
In the following, we will discuss the properties of the collective modes based on these expressions derived in this section. The discussions and the numerical calculations of the behaviors of the collective modes beyond the low-$q$ regime will be presented together.

\begin{figure}[!t]
\centering
\includegraphics[width=3.2in]{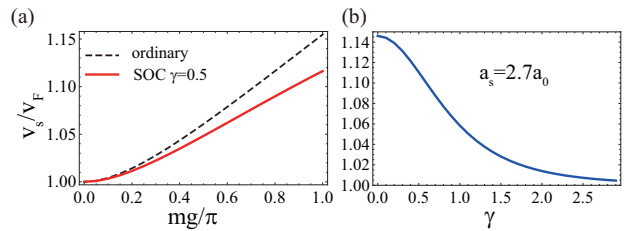}\hspace{0.5cm}
\caption{(Color online) (a) The speed of zero sound as a function of $s$-wave scattering length. (b) The speed of zero sound as a function of the strength of SOC. The parameters used here are: the number of $^{40}$K atoms is about $10^{4}$, $k_R=2\pi/\lambda$ with $\lambda=773$nm, $\gamma=0.5$, trapping frequencies $(\omega_{\bot},\omega_z)= 2 \pi \times (10, 400)$Hz, and $a_s=2.70 a_0$, where $a_0$ is the Bohr Radius. The corresponding dimensionless interaction strength $mg$ is about $3.0$, which is less than the critical value $\pi$.}
\label{fig:sound}
\end{figure}

\section{Collective modes}

In this section, we study the collective behaviors via the poles of the density and spin RPA correlation functions \cite{Negele}. At first, we discuss the low-$q$ behaviors based on the expressions presented in the last section, and then, we will show the dispersion relations of collective modes beyond low-$q$ regime through numerical calculations.

To study the low-$q$ behaviors of the collective modes, it is useful to note the following behaviors of $\chi^{\mu \nu}$. First, if the dimensionless ratio $y=m\omega /\kappa k_F\left\vert \mathbf{q}\right\vert$ is fixed when $q\rightarrow 0$, we obtain
\begin{equation}  \label{chi0:2}
\chi _{nT}=\left(
\begin{array}{cc}
\chi_{nn} & \frac{\gamma }{\kappa }y\chi_{nn} \\
\frac{\gamma }{\kappa }y\chi_{nn} & \frac{m}{\pi}+y^2\chi_{nn}
\end{array}
\right),\ \ |\mathbf{q}|\rightarrow 0, y \text{  fixed},
\end{equation}
where $\chi_{nn}=-mI_1(y)/\pi$ is the density-density element of the susceptibility matrix.
%Similar forms could be found in graphene sheets \cite{Principi}.
Secondly, if we fix the $\omega$ when $|\mathbf{q}|\rightarrow 0$, the dimensionless parameter $y$ tends to infinity. At first, we consider weak SOC case ($\gamma \ll 1$), and we obtain
\begin{equation}  \label{chi0:1}
\chi _{nT}=\frac{m}{\pi }\left(
\begin{array}{cc}
0 & 0 \\
0 & \frac{8\gamma ^{2} \epsilon_F^2}{16\gamma ^{2} \epsilon_F^2-\omega
^{2}}
\end{array}
\right), \ \ |\mathbf{q}|\rightarrow 0, \omega \text{  fixed},
\end{equation}
where the intraband has no contribution in this limit. From Eq. (\ref{chi0:2}) and (\ref{chi0:1}), we conclude that the solutions contain two different categories: (i) the gapless modes, which are coupled oscillations of the density and transverse spin excitations; (ii) the gapped modes, which are the oscillations of the transverse spin excitations.

%%%%%%%%%%%%%%%%%%%%%%%%%%%%%%%%%%%%%%%%%%%%%%%%%%%%%%%%%%%%%%%%%%%%%%%%%%%%%%%%%%%%%%%%%%%%%%%%%
% give another graph, show the energy gaps of T,L,Z by solving ....

\begin{figure}[!t]
\begin{center}
\includegraphics[width=3.2in]{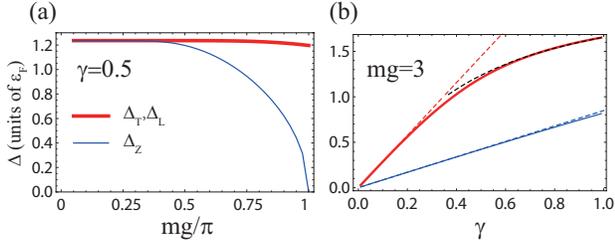}\hspace{0.5cm}
\end{center}
\caption{ (Color online) The energy gaps for gapped modes as functions of dimensionless interaction strength $mg$ in (a), and SOC strength $\gamma$ in (b). The parameters taken here are the same with Fig. \ref{fig:sound}. For (a), we find the energy gaps are close to the edge of particle-hole continuum for $mg/\pi < 0.5$. For (b), the red and blue dashed lines starting from $\gamma=0$ are approximations in Eqs. (\ref{gap1}) and (\ref{gap23}), and the black dashed line starting from $\gamma=1$ are boundary of the particle-hole continuum at $\mathbf{q}=0$.}
\label{fig:gaps}
\end{figure}

The collective modes in this subspace can be determined by the poles of RPA susceptibility in Eq. (\ref{eq:chiRPA}), which are given by
\begin{equation}\label{eq:det}
\text{det}(1+g\sigma_z \chi_{nT})=0.
\end{equation}
The solutions of the real part of Eq. (\ref{eq:det}) give rise to the dispersions of the collective modes. Within the scheme of RPA, we find that the determinant $\text{det}(1+g\sigma_z \chi_{nT})$ will develop a non-zero imaginary part when the dispersions fall in the particle-hole continuum, which is given in Fig. \ref{figure:phc}. For this regime, the collective modes are unstable and will decay to particle-hole pairs.

Substitute Eq. (\ref{chi0:2}) to Eq. (\ref{eq:det}), we obtain a gapless dispersion, which is given by the solution of
\begin{equation} \label{eq:zerosound}
(1+g\chi_{nn})[1-g(\frac{m}{\pi}+ y^2 \chi_{nn})]+\frac{\gamma^2}{\kappa^2}g^2 y^2 \chi_{nn}^2=0.
\end{equation}
From Eq. (\ref{chi0:2}), we find that the density component is coupled with the transverse spin component. Therefore the zero sound is a coupled oscillation of the density and transverse spin excitations. In Fig. \ref{fig:sound} (a), we show the sound speed of this gapless mode by solving Eq. (\ref{eq:zerosound}) numerically. For comparison, the sound speed of the ordinary Fermi liquid without SOC is shown together.

Substitute Eq. (\ref{chi0:1}) to Eq. (\ref{eq:det}), we obtain one branch of gapped mode corresponding to transverse spin oscillation. The frequency at $|\mathbf{q}|=0$ of this mode is given by the solution of
\begin{equation}
1-\frac{mg}{\pi}\frac{8\gamma ^{2} \epsilon_F^2}{16\gamma ^{2} \epsilon_F^2-\omega
^{2}}=0,
\end{equation}
which yields the result
\begin{equation} \label{gap1}
\Delta_T=4\sqrt{1-m g/2\pi } \gamma \varepsilon _{F},
\end{equation}
where $\Delta_T$ is the energy gap of the transverse spin mode, $mg=\varepsilon_{int}/\varepsilon_{kin}$ is the dimensionless ratio of interaction energy and kinetic energy in 2D, which denotes the significance of interaction strength in ultracold atomic system. The joint orbital and spin rotation about the $\hat{z}$ axis $[U(1)_{orbital}\times U(1)_{spin}]_D$ indicates that the low-q dispersion of this gapped mode as $\omega_T(\mathbf{q})=\Delta_T+\alpha_T q^2$.

Similarly, we consider susceptibility in the $s^L$-$s^Z$ subspace in the long wavelength limit under the assumption of weak SOC ($\gamma \ll 1$). The RPA susceptibility $\chi^{RPA} _{LZ}$ is positive definite in the limit $q \rightarrow 0$ while keeping $y$ fixed in the normal state regime. Therefore there are no gapless excitations in the the longitudinal and perpendicular spin subspace. In the limit $q \rightarrow 0$ while keeping $\omega$ fixed, the interband susceptibility dominates and gives rise to
\begin{equation}
\chi _{LZ}=\frac{m}{\pi }\left(
\begin{array}{cc}
\frac{8\gamma ^{2} \epsilon_F^2}{16\gamma ^{2} \epsilon_F^2-\omega
^{2}} & 0
\\
0 & \frac{16\gamma ^{2} \epsilon_F^2}{16\gamma ^{2} \epsilon_F^2-\omega
^{2}}
\end{array}%
\right).
\end{equation}
The poles of RPA susceptibility $\chi_{LZ}^{RPA}=\chi_{LZ}[1-g\chi_{LZ}]^{-1}$ give rise to two branches of gapped modes corresponding to the longitudinal and perpendicular spin oscillations. The frequencies at $|\mathbf{q}|=0$ of these two modes are given as follows
\begin{subequations}\label{gap23}
\begin{eqnarray}
\Delta_L&=4\sqrt{1-mg/2\pi }\gamma \varepsilon _{F},\\
\Delta_Z&=4\sqrt{1-mg/\pi }\gamma \varepsilon_{F},
\end{eqnarray}
\end{subequations}
where $\Delta_L$ and $\Delta_Z$ are the energy gaps for the longitudinal spin and perpendicular spin modes respectively. The joint orbital and spin rotation about the $\hat{z}$ axis $[U(1)_{orbital}\times U(1)_{spin}]_D$ indicates that the low-q dispersions of the two gapped modes are $\omega_L(\mathbf{q})=\Delta_L+\alpha_L q^2$ and $\omega_Z(\mathbf{a})=\Delta_Z+\alpha_Z q^2$. Compared with the formula of $\Delta_T$ in Eq. (\ref{gap1}), we notice that the energy gaps for transverse spin and longitudinal spin modes are degenerate: $\Delta_T=\Delta_L$. But the coefficients of $q^2$ terms need not to be the same. One can also see that above some g values, the energy gaps will vanish. However, the system may develop some magnetic orders before these g values \cite{SSZhang}

This degeneracy of $\Delta_T$ and $\Delta_L$ could be understood generally as follows. For finite momentum of the collective excitations $\mathbf{q}$, the in-plane spin could be divided into transverse ($s^T$) and longitudinal ($s^L$) components relative to the direction of $\mathbf{q}$ as defined in Sec. II. In the limit $\mathbf{q} \rightarrow 0$, the in-plane spin components $s^T$ and $s^L$ are not distinguishable. And therefore the susceptibility for the in-plane spin fluctuations are isotropic. As a result the corresponding energy gaps $\Delta_T$ and $\Delta_L$ equal to each other. For non-zero momentum, this degeneracy is lifted and the collective modes for $s^T$ and $s^L$ fluctuations have different dispersions, therefore, $\alpha_T\neq \alpha_L$ in general. This argument also applies to the case for strong SOC, which is shown explicitly in the following discussions.

\begin{figure}[!t]
\begin{center}
\includegraphics[width=2.3in]{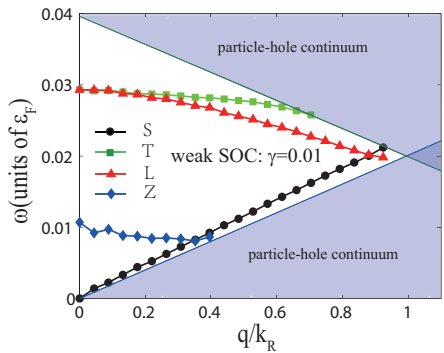}\hspace{0.5cm}
\end{center}
\caption{ (Color online) The collective excitations for weak SOC with $\gamma=0.01$. The other parameters used here are the same with Fig. \ref{fig:sound}. The transverse, longitudinal and perpendicular spin excitations are labeled by $T,L$ and $Z$. S denotes the zero sound mode. These collective modes finally disappear in the particle-hole continuum.}
\label{smallSOC}
\end{figure}

We now want to consider the collective modes for strong SOC. In the present experiments \cite{JZhang,Zwierlein}, the typical experimental values for the dimensionless ratio $\gamma$ ranges from about $0.5$ to $1$. Thus in the following, we need to consider the case for $\gamma \sim \mathcal{O}(1)$ based on Eq. (\ref{intraband}) for the gapless mode and Eq. (\ref{interband12}) for the gapped modes. The sound speed of the gapless mode with strong SOC has been shown in Fig. \ref{fig:sound} (b). We found the sound speed relative to the Fermi velocity is reduced with increasing SOC.

%In fact, Eq. (\ref{chi0:2}) holds both for weak and strong SOC in the limit $|\mathbf{q}|\rightarrow 0$ with $m\omega/\kappa k_F|\mathbf{q}|$ kept fixed, which originates from the usual continuity equation and has been explicitly calculated above. Similar forms could be found in graphene sheets \cite{Principi}.

There are still three gapped modes corresponding to transverse, longitudinal and perpendicular spin excitation and the analytical formulas for the energy gaps of gapped modes given by (\ref{gap1}) and (\ref{gap23}) provide qualitative approximations, with corrections $\mathcal{O}(\gamma^2)$. By numerically solving the RPA equation based on the exact formula in Eq. (\ref{intraband}), we obtain the energy gaps as a function of $mg$ and $\gamma$ in Fig. \ref{fig:gaps} for typical experimental parameters. From Fig. \ref{fig:gaps} (a), we find that the energy gaps are close to the boundary of the particle-hole continuum at $\mathbf{q}=0$ for $mg/\pi<0.5$, which is on the same order of $\gamma$. From Fig. \ref{fig:gaps} (b), we find in the weak SOC regime, the analytical results for the two in-plane modes T, L (shown as dashed lines) agree well with the numerical solutions. In the strong SOC regime with $\gamma \sim 1$, the energy gaps for the in-plane modes T, L are close to boundary of the particle-hole continuum (shown as the black dashed line). The analytical result for out-plane modes Z (shown as blue dashed line) shows a well agreement with the numerical solution in the full range of $\gamma$.

\begin{figure}[!t]
\begin{center}
\includegraphics[width=2.3in]{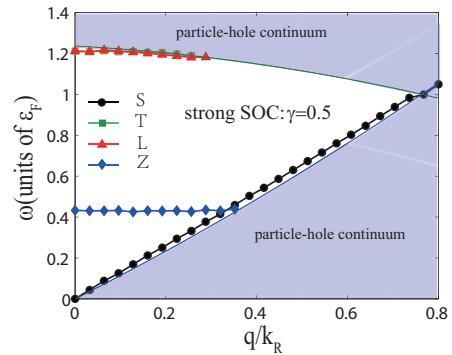}\hspace{0.5cm}
\end{center}
\caption{ (Color online) The collective excitations for strong SOC with $\gamma=0.5$. The other parameters and the notations of these collective modes are the same with Fig. \ref{smallSOC}.}
\label{largeSOC}
\end{figure}

Finally, we want to stress that the cold atom systems are prepared in harmonic traps. Although the trap size is usually much larger than the interatom distance and the laser wave length \cite{Bloch}, the finite size effect of the trap should be concerned. Results obtained from uniform system in this work could be applied to the trapped system only when the wave lengths of the collective excitations are much smaller than the trap size $a_{\text{HO}}$. More specifically, the finite size of the system provides a lower limit $2\pi/a_{\text{HO}}$ for the momentum scale of the collective modes. To go beyond the low-$q$ regime, we numerically evaluated the dispersion relations $\omega_i(\mathbf{q}),i=S,T,L,Z$ shown in Figs. \ref{smallSOC} and \ref{largeSOC}, corresponding to weak and strong SOC respectively. At finite momentum, the gapped mode $\omega_T(\mathbf{q})$ becomes a coupled oscillation of density and transverse spin due to SOC. For SOC generated in present experiments \cite{JZhang,Zwierlein}, $\gamma \sim 1$, the dispersion relations for modes T and L are nearly degenerate and stand close to the edge of the particle-hole continuum (see Fig. \ref{largeSOC}). The dispersion relations for gapped modes are more flatter compared to the weak SOC case.

\section{experimental signatures and summaries}

We have shown the behaviors of collective modes in the long wave length limit with SOC and repulsive $s$-wave interaction. Recently, the SOC in Fermi gas has been realized wtih $^{40}$K atoms \cite{JZhang} and $^{6}$Li atoms \cite{Zwierlein}. In their experiments, the equal weight combination of Rashba-type and Dresselhaus-type SOC is realized. This is the first step towards the realization of pure type of SOC experimentally. In this work, we focus on the Rashba-type SOC, and the Dresselhaus-type SOC is presented in Appendix B, which is demonstrated to give the same results as the Rashba SOC for collective behaviors. The short-ranged repulsive $s$-wave interaction can be achieved on the upper branch of a Feshbach resonance, where there are uncondensed Fermi gases in the absence of molecule formation \cite{TLHo}. The repulsive Fermi gas is metastable for observation when it is far away from the resonant regime, and has been successfully reached in the recent experiment \cite{GBJo}.

We choose the following typical experimental parameters for quasi-2D system considered here. We consider about $10^4$ $^{40}$K atoms in a pancake-shaped harmonic potential with the trapping frequencies $2\pi \times(10,10,400)$Hz along the $(\mathbf{\hat{x}},\mathbf{\hat{y}},\mathbf{\hat{z}})$ direction. The system size is estimated as $(37.8,37.8,5.98)\mu$m. The strength of SOC $\gamma$ is chosen as $0.5$ and $a_s$ is tuned to $2.7 a_0$ (within the normal state regime) with Feshbach resonance. For the parameters used here, we estimate that the zero sound velocity is about $1.1v_F$, and the energy gaps for the gapped modes are $\Delta_T=\Delta_L=1.21\epsilon_F$ and $\Delta_Z=0.42\epsilon_F$. The Fermi velocity $v_F$ is about $0.028$m/s and the Fermi energy $\epsilon_F=h \times 33.36$kHz. Furthermore, the realistic system is prepared in a trap. The results for a uniform system can be used only if the wavelength of the excitation is much smaller than the trap size. To go beyond the low-$q$ regime, we show the dispersion relations of the collective modes in Fig. \ref{largeSOC} for typical experimental parameters. To observe these dynamical oscillations in experiment, we suggest the following methods. (i): The gapless mode could be excited by a short laser pulse focused near the center of the trap \cite{Recati}. The oscillation could be detected via the spatially resolved images of the coupled density and transverse spin perturbances propagating through the trapped atomic cloud \cite{McGuirk}. (ii): The gapped modes are spin oscillations, which are actually the oscillations of the internal hyperfine states of atoms. The oscillations could be excited via a two-photon drive, and traced out through the state-selective absorption imaging method and repeating the experiment for many values of evolution time \cite{McGuirk,Lewandowski}.

In summary, we studied the collective modes of spin-orbit coupled Fermi gas with repulsive $s$-wave interaction. There are two categories of collective modes in this system. One branch has a gapless dispersion, known as the zero sound. In presence of SOC, the density oscillation is intrinsically coupled with the transverse spin oscillation for this mode. The other three branches are collective excitations with finite energy gaps, which are closely related to the energy split due to SOC. We calculated the sound speed of the gapless mode and the energy gaps of the gapped modes, and also estimated their values for typical experimental parameters. In contrast to the Coulomb interaction in 2D solid state systems, the $s$-wave interaction leads to fundamentally different phenomena, such as the presence of linear dispersion of the zero sound and the gapped modes. The study on the collective modes of the SOC repulsive Fermi gas indicates some novel behaviors due to the presence of SOC, and also might have the immediate applicability to experimental study of the SOC Fermi gases in the upper branch of the energy spectrum.

\bigskip

\begin{acknowledgments}
We acknowledge helpful discussions with Han Pu, Congjun Wu, and G. Juzeli$\bar{\text{u}}$nas.
This work was supported by the NKBRSFC under grants Nos. 2011CB921502, 2012CB821305, 2009CB930701, 2010CB922904, NSFC under grants Nos. 10934010, 11228409, 61227902 and NSFC-RGC under grants Nos. 11061160490 and 1386-N-HKU748/10. J. Ye was supported by NSF-DMR-1161497, NSFC-11074173, 11174210, Beijing Municipal Commission of Education under Grant No. PHR201107121, at KITP was supported in part by the NSF under grant No. PHY11-25915.
\end{acknowledgments}

\appendix

\section{EVALUATIONS OF SOME RELEVANT INTEGRALS}

When we calculated the density and spin susceptibility in Sec. III, we met with some integrals of
the azimuthal angle $\theta $, such as $I_{0},I_{1},I_{2}, I_{3},
I_{4},I_{5}$. Since they have the similar structures, we evaluate
them in this Appendix. The definitions of these integrations are listed as
bellow:
\begin{eqnarray}
I_{0}\left( y\right) &=&\int \frac{d\theta }{2\pi }\frac{1}{y-\cos \theta
+i0^{+}},  \label{eq:integrand1} \\
I_{1}\left( y\right) &=&\int \frac{d\theta }{2\pi }\frac{\cos \theta }{%
y-\cos \theta +i0^{+}},  \label{eq:integrand2} \\
I_{2}\left( y\right) &=&\int \frac{d\theta }{2\pi }\frac{\cos ^{2}\theta }{%
y-\cos \theta +i0^{+}},  \label{eq:integrand3} \\
I_{3}\left( y\right) &=&\int \frac{d\theta }{2\pi }\frac{\cos ^{3}\theta }{%
y-\cos \theta +i0^{+}},  \label{eq:integrand4} \\
I_{4}\left( y\right) &=&\int \frac{d\theta }{2\pi }\frac{\sin ^{2}\theta }{%
y-\cos \theta +i0^{+}},  \label{eq:integrand5} \\
I_{5}\left( y\right) &=&\int \frac{d\theta }{2\pi }\frac{\cos \theta \sin
^{2}\theta }{y-\cos \theta +i0^{+}}.  \label{eq:integrand6}
\end{eqnarray}%
At first we note that if $\left\vert y\right\vert <1$, these integrals have a non-zero imaginary part; if $\left\vert y\right\vert >1$, these integrals are real. This can be seen from the following calculation. This property results in the damping of the collective modes in the particle-hole excitation continuum.

\begin{figure}[!h]
\begin{center}
\includegraphics[width=3in]{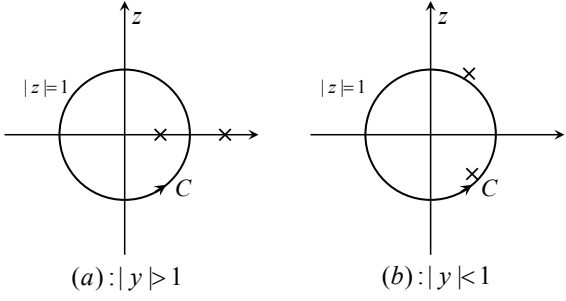}\hspace{0.5cm}
\end{center}
\caption{ Schematic of the integral path and the poles of
the integrand functions. The location of the poles are different for $|y|>1$
and $|y|<1$, which are shown in (a) and (b) respectively.}
\label{fig:pathC}
\end{figure}

These integrals can be mapped into the integrals in the complex-$z$ plane with a transformation $z=e^{i\theta }$. The integral path $\mathcal{C}$ is the unit circle with the center at $(0,0)$. We show the integral path and gives a schematic of the poles of the integrand function for $|y|>1$ and $|y|<1$ in Fig. \ref{fig:pathC}. With the theorem of residue, these integrals are evaluated straightforwardly:
%\begin{widetext}
\begin{eqnarray}
I_{0}\left( y\right)  &=&\frac{\text{sign}(y)}{\sqrt{y^{2}-1}}\Theta \left(
|y| -1\right) \notag \\
&& -\frac{i}{\sqrt{1-y^{2}}}\Theta(
1-|y|), \label{eq:intresult1}\\
I_{1}\left( y\right)  &=&-1+\frac{|y|}{\sqrt{y^{2}-1}}\Theta(|y| -1) \notag \\
&&  -\frac{i y}{\sqrt{1-y^{2}}}\Theta (
1-|y|), \label{eq:intresult2}\\
I_{2}(y) &=&-y+\frac{y\left\vert y\right\vert }{\sqrt{y^{2}-1}}\Theta
\left( \left\vert y\right\vert -1\right) \notag \\
&&  -i\frac{y^{2}}{\sqrt{1-y^{2}}}
\Theta ( 1-|y|), \label{eq:intresult3}\\
I_{3}(y) &=&-\frac{1}{2}-y^{2}+\frac{y^{2}\left\vert y\right\vert }{\sqrt{%
y^{2}-1}}\Theta \left( \left\vert y\right\vert -1\right) \notag \\
&& -\frac{iy^{3}}{%
\sqrt{1-y^{2}}}\Theta \left( 1-\left\vert y\right\vert \right) , \label{eq:intresult4}\\
I_{4}(y) &=&y-\text{sign}(y)\sqrt{y^{2}-1}\theta (|y|-1) \notag \\
&&-i\sqrt{1-y^{2}}\theta
(1-|y|)], \label{eq:intresult5}\\
I_{5}(y) &=&-\frac{1}{2}+y^{2}-\left\vert y\right\vert \sqrt{y^{2}-1}%
\Theta \left( \left\vert y\right\vert -1\right) \notag \\
&&-iy\sqrt{1-y^{2}}\Theta
\left( 1-\left\vert y\right\vert \right).\label{eq:intresult6}
\end{eqnarray}
%\end{widetext}
where $\Theta (x)$ is the unitstep function. When we were exploring the low-$q$ behaviors of gapped modes in the region $q\ll k_{R}$, we used the asymptotic behaviors at $|y|\rightarrow \infty $:
\begin{eqnarray}  \label{eq:asymptotic}
I_{0}\left( y\right) &\simeq &\frac{1}{y},I_{1}\left( y\right) \simeq \frac{%
1}{2y^{2}},I_{2}\left( y\right) \simeq \frac{1}{2y},  \notag \\
I_{3}\left( y\right) &\simeq &\frac{3}{8y^{2}},I_{4}\left( y\right) \simeq
\frac{1}{2y},I_{5}(y)\simeq \frac{1}{8y^{2}}.
\end{eqnarray}

\bigskip

\section{EQUIVALENCE BETWEEN THE RASHBA SOC AND DRESSELHAUS SOC}

We start with the single particle Hamiltonian wtih Dresselhaus SOC
\cite{NIST4}%
\begin{equation}
\mathcal{H}_{D}=\frac{\mathbf{k}^{2}}{2m}+\alpha(-k_{y}\sigma _{x}-k_{x}\sigma _{y})-\mu.
\end{equation}%
The various quantities calculated in this paper is closely based on the well-defined
Feynman rules. The Feynman rules include the single particle Green's
function and the interaction vertex. Within the scheme of RPA, the collective modes are determined by the poles of the RPA susceptibility, which is given by (\ref{eq:chiRPA}). It is apparent
that the properties of the collective modes depends on the bare susceptibility $\chi\left( \mathbf{k},\omega \right) $. Therefore, we want to derive the relationships between the two types of SOC system as follows.

For the Dresselhaus-type SOC, the non-interacting Green's function and the
interaction vertex reads
\begin{eqnarray}
G^{D}\left( \omega ,\xi _{\mathbf{k,}s}\right) &=&\sum\limits_{s}\frac{%
P_{s}^{D}}{\omega -\xi _{\mathbf{k,}s}+i0^{+}}, \\
V_{ss^{\prime };rr^{\prime }}^{D}(\mathbf{k,p,q}) &=&gf_{ss^{^{\prime
}}}^{D}(\theta _{\mathbf{k}},\theta _{\mathbf{k+q}})f_{rr^{^{\prime
}}}^{D}(\theta _{\mathbf{p}},\theta _{\mathbf{p-q}}).
\end{eqnarray}%
where the upscript $D$ represents the Dresselhaus-type SOC, and $P^{D}_{s}$
is the projection operator defined as
\begin{equation}\label{projectionD}
P_{s}^{D}=\frac{1}{2}[1+s(-k_{y}\sigma _{x}-k_{x}\sigma _{y})].
\end{equation}
The energy spectrum of the Dresselhaus-type SOC is the same with the Rashba-type. However, the spin polarization is different from Rashba SOC, which is
\begin{equation}\label{projectionR}
P_{s}^{R}=\frac{1}{2}[1+s(-k_{y}\sigma _{x}+k_{x}\sigma _{y})].
\end{equation}

The trace in Eq. (\ref{chi0}) includes an overlap factor for the Dresselhaus case as
\begin{equation}
F_{sr}^{D,\mu \nu }=tr\left[ P_{s}^{D}\left( \mathbf{k-q/2}\right) \sigma
_{\mu }P_{r}^{D}\left( \mathbf{k+q/2}\right) \sigma _{\nu }\right] .
\end{equation}%
Given Eq. (\ref{projectionD}) and (\ref{projectionR}), the overlap factor for the two cases can be related by transformation $k_{x}\rightarrow \mathbf{-}k_{x},q_{x}\rightarrow \mathbf{-}q_{x}$, while keeping the $y$ component invariant. The bare susceptibility with Dresselhaus SOC is given by
\begin{equation}
\chi^{D,\mu \nu }\left( \mathbf{q},\omega \right) =-\sum\limits_{%
\mathbf{k,}s,r}F_{sr}^{D,\mu \nu }\frac{f\left( \xi _{\mathbf{k-q/}%
2,s}\right) -f\left( \xi _{\mathbf{k+q/}2,r}\right) }{\xi _{\mathbf{k-q/}%
2,s}-\xi _{\mathbf{k}+\mathbf{q}/2,r}+\omega +i0^{+}}.
\end{equation}%
Due to the rotation symmetry, we choose $\mathbf{q}=q\mathbf{e}_{y}$ for simplicity. Given the relationship between the overlap factor for Rashba SOC and Dresselhaus SOC and the rotational symmetry of the energy spectrum, we have
\begin{equation}
\chi^{D,\mu \nu }\left( q\mathbf{e}_{y},\omega \right) =\chi^{R,\mu \nu }\left( q\mathbf{e}_{y},\omega \right) .
\end{equation}

Therefore we conclude that all the properties and classification of the
collective modes for the two type of SOC are all the same exactly.

\end{document}